\documentclass[aps,showpacs,superscriptaddress]{revtex4}
\usepackage{times} 
\usepackage{amssymb}
\usepackage{amsbsy}
\usepackage{amsmath}
\usepackage{amsbsy}
\usepackage{latexsym}
\usepackage[usenames,dvipsnames]{xcolor}

\begin{document}

\title{Electronic Energy Functionals: Levy-Lieb principle within the Ground State Path Integral Quantum Monte Carlo}
\author{Luigi Delle Site}
\affiliation{Institute for Mathematics, Freie Universit\"{a}t Berlin,
Arnimallee 6, D-14195 Berlin, Germany}
\email{luigi.dellesite@fu-berlin.de}
\author{Luca M.Ghiringhelli}
\affiliation{Fritz-Haber-Institut, Faradayweg 4--6, D-14195 Berlin-Dahlem,
  Germany}
\email{ghiringhelli@fhi-berlin.mpg.de}
\author{David M.Ceperley}
\affiliation{Department of Physics and NCSA, University of Illinois at Urbana-Champaign, Urbana, Illinois 61801, USA}
\email{ceperley@ncsa.uiuc.edu}
\begin{abstract}
We propose a theoretical/computational protocol based on the use of the Ground State (GS) Path Integral (PI) Quantum Monte
Carlo (QMC) for the calculation of the kinetic and Coulomb energy density
for a system of $N$ interacting electrons in an external potential. The idea
is based on the derivation of the energy densities via the $N-1$-conditional
probability density within the framework of the Levy-Lieb constrained search
principle. The consequences for the development of energy functionals within
the  context of Density Functional Theory (DFT) are discussed.
We propose also the possibility of going beyond the energy
densities and extend this idea to a computational procedure where the
$N-1$-conditional probability is an implicit functional of the electron
density, independently from the external potential. In principle, such a
procedure paves the way for an {\it on-the-fly} determination of the energy
functional for any system.

\end{abstract}
\maketitle
\section{Introduction}
\subsubsection{Levy-Lieb Constrained principle} 
M.Levy and E.Lieb \cite{levy,lieb} have, independently from each other,
provided a general minimization principle which leads to the rigorous
definition of the universal functional of Hohenberg and Kohn in Density
Functional Theory (DFT)\cite{hk,py}. The equation for the ground state energy in the Levy-Lieb (LL) formulation is:
\begin{equation}
E_{GS}=\min_{\rho}\left[\min_{\psi\rightarrow\rho}\left<\psi|K+V_{ee}|\psi\right>+\int
  \rho({\bf r})v({\bf r})d{\bf r}\right].
\label{ll1} 
\end{equation}
with $E_{GS}$ the ground state energy, $K$ the kinetic and $V_{ee}$ the electron-electron Coulomb
operator, $\rho({\bf r})$ the one-particle electron density and $v({\bf r})$ the external potential (e.g., electron-nucleus Coulomb interaction).
The meaning of Eq.\ref{ll1} is that the minimization over $\psi$ is restricted to all antisymmetric
wavefunctions such that $\rho({\bf r})=N\int\psi^{*}({\bf r}, {\bf r}_{2},.....{\bf r}_{N})\psi({\bf r}, {\bf r}_{2},.....{\bf r}_{N}) d{\bf r}_{2}...d{\bf r}_{N}$, while the outer
minimization searches over all the $\rho$'s which integrate to $N$, number of particles.
The rigorous definition of the universal functional of Hohenberg and Kohn follows as:
\begin{equation}
F[\rho]=\min_{\psi\rightarrow\rho}\left<\psi|K+V_{ee}|\psi\right>.
\label{llf}
\end{equation}
Obviously, searching on the whole space of antisymmetric
wavefunctions is possible only in abstract terms and becomes impossible
when one tries to actually apply the LL principle and derive an explicit
expression of the universal functional as a functional of $\rho({\bf r})$.  In
order to circumvent this difficulty and make it possible the derivation of a
functional, one would need a formalism which expresses
Eq.\ref{llf} in terms of $\rho$, removing the explicit dependence on $\psi$; such a formalism is  reported below.
\subsubsection{The Levy-Lieb principle in terms of the $(N-1)$ conditional probability density}
Let us consider the properly normalized $3N$-dimensional probability density of an $N$-electron system:
\begin{equation}
N\psi^{*}({\bf r},{\bf r}_{2}....{\bf r}_{N})\psi({\bf r},{\bf r}_{2}....{\bf
  r}_{N})=\Theta({\bf r},{\bf r}_{2}........{\bf r}_{N})
\label{teta}
\end{equation}
this can be equivalently written as: 
\begin{equation}
\Theta({\bf r},{\bf r}_{2}....{\bf r}_{N})=\rho({\bf r})f({\bf r}_{2},....{\bf
  r}_{N}|{\bf r})
\label{fact}
\end{equation}
where $\rho({\bf r})$ is the one particle electron density
(normalized to $N$) and $f({\bf r}_{2},....{\bf
  r}_{N}|{\bf r})$ is the $(N-1)$ electron conditional (w.r.t. ${\bf
  r}$) probability density. This latter in nothing else than the probability density of finding a
configuration of $(N-1)$ particles after the position of one specific particle has been fixed \cite{dinur,note}. In order to write the density functional in the standard notation used in literature, here we have identified ${\bf r}_{1}$ with ${\bf r}$.
The LL principle of Eq.\ref{ll1} can then be rewritten as \cite{dinur,mio2,mio3}: 
\begin{equation}
E_{GS}=\min_{\rho}\left[\left(\min_{f}\Gamma[\rho,f]\right)+\frac{1}{8}\int\frac{|\nabla\rho({\bf
      r})|^{2}}{\rho({\bf r})}d{\bf r}+\int \rho({\bf r})v({\bf r}) d{\bf
    r}\right]
\label{eqg1}
\end{equation}
with
\begin{equation}
F[\rho]=\left(\min_{f}\Gamma[\rho,f]\right)+\frac{1}{8}\int\frac{|\nabla\rho({\bf
      r})|^{2}}{\rho({\bf r})}d{\bf r}
\label{eqg2}
\end{equation}
and
\begin{eqnarray}
\Gamma[\rho,f]=\frac{1}{8}\int \rho({\bf
  r})\left[\int_{{\bf R}^{N-1}}\frac{|\nabla_{{\bf r}}f({\bf r}_{2},....{\bf
  r}_{N}|{\bf r})|^{2}}{f({\bf r}_{2},....{\bf
  r}_{N}|{\bf r})}d{\bf r}_{2}....d{\bf r}_{N}\right]d{\bf r}+\\+(N-1)\int \rho({\bf
  r})\left[\int_{{\bf R}^{N-1}}\frac{f({\bf r}_{2},....{\bf
  r}_{N}|{\bf r})}{|{\bf r}-{\bf r}_{2}|}d{\bf r}_{2}....d{\bf r}_{N}\right]d{\bf r}.
\label{gamma}
\end{eqnarray}
Where ${\bf R}^{N-1}$ denotes the space of configuration $({\bf
  r}_{2},..........{\bf r}_{N})$.
The inner minimization searches for the $f$ which minimizes
$\Gamma[\rho,f];$~$\forall \rho$. Here we underline the fact that the above formalism does not contain approximations, i.e. the ground state identified in Eq. \ref{eqg1} is the same which solves the time-independent Schr\"{o}dinger equation with the same Hamiltonian.\\
The central question, is how to determine $f$ in an efficient way and once there is a procedure for doing so, how this can be used in concrete terms within the DFT framework. In our previous work \cite{luca1,luca2}, we adopted a physically-motivated explicit guess functional form for $f$, dependent on one free parameter, and we numerically optimized the resulting $\Gamma[\rho,f]$ w.r.t. the single parameter. Here, we propose a radical step further, by leaving the functional form of $f$ completely undetermined and (numerically) derive it within an \textit{exact} quantum Monte Carlo framework. 
In the following part of this work we suggest two different but related methodologies, a) one related to the calculation of the energy density of the ground state which can then be used as a reference for developing analytic functionals 
and b) another where $f$ can be determined as a numerical functional of $\rho$, independently of the external potential, and thus provide a {\bf numerically} exact route to the calculation of the universal Hohenberg-Kohn functional.
It must be taken into account that the intention of this paper is to provide  a theoretical/methodological guideline and its practical warnings; at this stage we do not provide numerical experiments. In fact, we hope that the optimal computational implementation of the approach will come from a constructive discussion of the ideas reported here.   
\section{Energy Density of the Ground State}
If we restrict ourselves to the ground state of a specific system of $N$
electrons with a well defined external potential, then the procedure of inner minimization of Eq.\ref{eqg1} (i.e. the  search for $f$ which minimizes
$\Gamma[\rho,f],~\forall \rho$) leads to $f_{\min}=f_{GS}$.\\
Let us define:
\begin{equation}
I({\bf r})=\int_{{\bf R}^{N-1}}\frac{|\nabla_{{\bf r}}f({\bf r}_{2},....{\bf
  r}_{N}|{\bf r})|^{2}}{f({\bf r}_{2},....{\bf
  r}_{N}|{\bf r})}d{\bf r}_{2}....d{\bf r}_{N}
\label{defi}
\end{equation}
and
\begin{equation}
C({\bf r})=\int_{{\bf R}^{N-1}}\frac{f({\bf r}_{2},....{\bf
  r}_{N}|{\bf r})}{|{\bf r}-{\bf r}_{2}|}d{\bf r}_{2}....d{\bf r}_{N}
\label{defc}
\end{equation}
since $f_{\min}=f_{GS}$, the explicit expression of the functional $F[\rho]$ is:
\begin{equation}
F[\rho]=\int\rho({\bf r})\left[\frac{1}{8}\frac{|\nabla\rho({\bf r})|^{2}}{\rho({\bf r})^{2}}+\frac{1}{8}I_{f_{GS}}({\bf r})+(N-1)C_{f_{GS}}({\bf r})\right]d{\bf r}.
\label{deff}
\end{equation}
The term:
\begin{equation}
\epsilon({\bf r})=\frac{1}{8}\frac{|\nabla\rho({\bf r})|^{2}}{\rho({\bf r})^{2}}+\frac{1}{8}I_{f_{GS}}({\bf r})+(N-1)\, C_{f_{GS}}({\bf r})
\label{deff1}
\end{equation}
is an energy density per particle expressed in terms of its kinetic $\left(\frac{1}{8}\frac{|\nabla\rho({\bf r})|^{2}}{\rho({\bf r})^{2}}+\frac{1}{8}I_{f_{GS}}({\bf r})\right)$ and Coulomb $\left((N-1)\, C_{f_{GS}}({\bf r})\right)$ parts.\\
Here, $I_{f_{GS}}({\bf r})$ and $C_{f_{GS}}({\bf r})$ indicate that the
quantities of Eq.\ref{defi} and Eq.\ref{defc} are those calculated for the $f$
of the ground state.\\
If one knew $I_{f_{GS}}({\bf r})$ and $C_{f_{GS}}({\bf r})$ as a functional of
$\rho$, this would correspond to have the universal functional of
Hohenberg and Kohn. However, even if $I_{f_{GS}}({\bf r})$ and
$C_{f_{GS}}({\bf r})$ are known only in a case-by-case situation, i.e., as functions of the
position (and not functional of the density), the expression of Eq.\ref{deff1}
would represent
the energy density of the universal functional, in the ground state of a chosen specific system.
This means that the energy density of any proposed functional in literature
should correspond to $\epsilon({\bf r})$ of Eq.\ref{deff1} when calculated in
the ground state.
At this point the key question is whether there is any rigorous technique
which can, in practical terms (i.e., not only formally), calculate the
$f_{GS}$ and thus determine $I_{f_{GS}}({\bf r})$ and
$C_{f_{GS}}({\bf r})$ \cite{mio3}.
If this is the case, then for any given system -- once the
number of particles is fixed -- one would have an explicit algorithm to compute
the numerically exact functional of $\rho$ in the ground state. In the next section, we propose that the
determination of the minimizing $f_{GS}$ 
and of the corresponding $I_{f_{GS}}({\bf r})$ and $C_{f_{GS}}({\bf r})$ 
can be achieved by the Ground State (GS) Path Integral (PI) Quantum Monte
Carlo (QMC) technique.

\section{Ground State Path Integral Quantum Monte Carlo for Fermions}
The Path Integral \cite{david1,david2} ground state \cite{sarsa} approach (GSPI) allows one to write the quantum partition function
of a system of $N$-particles as:
\begin{equation}
Z=\int d{\bf R}_{0}.....d{\bf R}_{M} \psi({\bf R}_{0})\exp[-S({\bf R}_{0}, {\bf
    R}_{1}....{\bf R}_{M})]\psi({\bf R}_{M})
\label{partfu}
\end{equation}
here ${\bf R}_{0}=({\bf r}^{0},{\bf r}^{0}_{2},.....{\bf r}^{0}_{N})$ is a
configuration of the $N$ particles in space, equivalently ${\bf R}_{1}$ is
another configuration and so on. In this way the sequence ${\bf
  R}_{0}.....{\bf R}_{M}$ represents an open path of length $M$ in the spaces of the
$N$-particle configurations. $\psi({\bf R}_{0})$ and $\psi({\bf R}_{M})$ is a
trial wavefunction calculated at the initial and final configuration. \\
Conceptually, the choice of $\psi$ is immaterial, since the evaluated quantities do not depend on it. However, technically, the choice of a good trial $\psi$ enhance the convergence of the method. 
$S({\bf R}_{0}, {\bf R}_{1}....{\bf R}_{M})$ is the action defined such that:
\begin{equation}
\exp[-S({\bf R}_{0}, {\bf R}_{1}....{\bf R}_{M})]=\left<{\bf R}_{0}|e^{-\tau H}|{\bf R}_{1}\right>\left<{\bf R}_{1}|e^{-\tau H}|{\bf R}_{2}\right>.....
\left<{\bf R}_{M-1}|e^{-\tau H}|{\bf R}_{M}\right>
\end{equation}
where $\tau$, which is formally an imaginary time, is: $\tau=\frac{\beta}{M}$, with $\beta$ formally the Boltzmann factor (the temperature has no physical meaning, but rather a parameter that influences the convergence efficiency of the method) and $H$ the Hamiltonian.\\ 
In this way, the quantum mechanical partition function is written as an integral involving a sequence of transitional probabilities in imaginary time $\tau$. Each of these transition probabilities can be decomposed into a kinetic part:
\begin{equation}
\left<{\bf R}_{i}|e^{-\tau K}|{\bf R}_{i+1}\right>=\frac{1}{(2\pi\tau)^{3N/2}}e^{-\frac{\tau}{2}\left(\frac{{\bf R}_{i}-{\bf R}_{i+1}}{\tau}\right)^{2}}
\label{spring}
\end{equation}
with $K$ being the kinetic operator, and a potential part:
\begin{equation}
\left<{\bf R}_{i}|e^{-\tau V}|{\bf R}_{i+1}\right>=\frac{1}{(2\pi\tau)^{3N/2}}e^{-\frac{\tau}{2}[V({\bf R}_{i})+V({\bf R}_{i+1})]}
\label{pot}
\end{equation}
with $V$ being the potential operator of the system considered.

\section{GSPI for electrons and calculation of $f$ and $\Gamma$}
In case of atoms and molecules, containing electrons (i.e. fermions) $V({\bf R})=V_{ee}+V_{ne}$, namely the electron-electron and the nucleus-electron interaction.
For fermions, in the case of a real $\psi({\bf R})$, one uses the fixed node
condition:
\begin{eqnarray}
V_{fermions}({\bf R})=V({\bf R})~~for~~~\psi({\bf R})>0 \\
V_{fermions}({\bf R})=\infty~~~~~~~for~~~\psi({\bf R})\le 0.
\end{eqnarray}
In case $\psi({\bf R})$ is complex, a term is added to the free-particle part
of the action. \\
Since the wavefunction 
is defined as: $\psi_{\tau}({\bf R})=e^{-\tau H}\Psi({\bf R})$,
where $\Psi({\bf R})$ is the ground state wavefunction, for $\tau$ that goes
to infinity, $\psi$ goes to the exact ground state wavefunction. Technically \cite{david1}, the wavefunction is evaluated at the midpoint of the path, i.e. at ${\bf R}_{M/2}$. 
In order to proceed in the derivation of $f$ and $\Gamma$ in terms of the GSPI approach, we adopt the following convention:
we will indicate the configuration at the midpoint of the path, ${\bf R}_{M/2}$ as ${\bf R}_{*}$. This means that $({\bf r}^{M/2},{\bf r}_{2}^{M/2},........{\bf r}_{N}^{M/2})$ becomes $({\bf r}^{*},{\bf r}^{*}_{2},........{\bf r}^{*}_{N})$.
 According to Eq.\ref{partfu}, the $N-1$-conditional probability density $f$
 can now be written as:
\begin{eqnarray}
f({\bf r}^{*}_{2},....{\bf r}^{*}_{N}|{\bf r}^{*})=\frac{1}{Z_{{\bf r}^{*}}}\int d{\bf R}_{0}d{\bf R}_{1}.......d{\bf R}_{\frac{M}{2}-1}d{\bf R}_{\frac{M}{2}+1}.......d{\bf R}_{M} \nonumber\\ \psi({\bf R}_{0})\exp[-S({\bf R}_{*}, {\bf
    R}_{0},{\bf R}_{1},....{\bf R}_{\frac{M}{2}-1},{\bf R}_{\frac{M}{2}+1}......{\bf R}_{M})]\psi({\bf R}_{M})
\label{effe1}
\end{eqnarray}
where:
\begin{eqnarray}
Z_{{\bf r}^{*}}=\int d{\bf R}_{*}^{N-1}d{\bf R}_{0}d{\bf R}_{1}....d{\bf R}_{\frac{M}{2}-1}d{\bf R}_{\frac{M}{2}+1}...d{\bf R}_{M} \psi({\bf R}_{0})\exp[-S({\bf R}_{*},{\bf R}_{0}, {\bf
    R}_{1},......{\bf R}_{\frac{M}{2}-1},{\bf R}_{\frac{M}{2}+1},....{\bf R}_{M})]\psi({\bf R}_{M})
\label{zetaf}
\end{eqnarray}
$d{\bf R}_{*}^{N-1}$ means that the integration is done on the whole space of
configurations ${\bf r}^{*}_{2},...{\bf r}^{*}_{N}$ of ${\bf R}_{*}$ except
that corresponding to variable ${\bf r}^{*}$.
With this set up, $f$ can be calculated by propagating
stochastically, according to a Monte Carlo procedure, the path ${\bf R}$ in imaginary time $\tau$.
Since the GSPI procedure, when evaluating in ${\bf R}_{*}$, delivers the ground state wavefunction of the system,
the expression of  $f$ in Eq.\ref{effe1} corresponds to the ground state
$N-1$-conditional probability density, 
that is, it corresponds to $f_{\min}$ of Eq.\ref{deff}.
The expression of Eq.\ref{effe1} can be introduced into Eq.\ref{defi}
and Eq.\ref{defc}; this leads to:
\begin{equation}
I({\bf r}^{*})=\int_{{\bf R}^{N-1}}\frac{|\nabla_{{\bf r}^{*}}f({\bf r}^{*}_{2},....{\bf
  r}^{*}_{N}|{\bf r}^{*})|^{2}}{f({\bf r}^{*}_{2},....{\bf
  r}^{*}_{N}|{\bf r}^{*})}d{\bf r}^{*}_{2}....d{\bf r}^{*}_{N}
\label{irho}
\end{equation}
and
\begin{equation}
C({\bf r}^{*})=\int_{{\bf R}_{N-1}}\frac{f({\bf r}^{*}_{2},....{\bf
  r}^{*}_{N}|{\bf r}^{*})}{|{\bf r}^{*}-{\bf r}^{*}_{2}|}d{\bf r}^{*}_{2}....d{\bf r}^{*}_{N}.
\label{crho}
\end{equation}
Where now $I_{f_{\min}}({\bf r})=I({\bf r}^{*})$ and $C_{f_{\min}}({\bf r})=C({\bf r}^{*})$.
Moreover, in $I({\bf r}^{*})$, the gradient, $\nabla_{{\bf r}^{*}}f({\bf
  r}^{*}_{2},....{\bf r}^{*}_{N}|{\bf r}^{*})$ can be calculated analytically
and thus sampled without additional computational costs in the MC sampling in configuration space; the explicit calculation is
reported in the Appendix.
The Hohenberg-Kohn functional in {\it local} form becomes:
\begin{equation}
F[\rho]=\frac{1}{8}\int\frac{|\nabla\rho({\bf
      r}^{*})|^{2}}{\rho({\bf r}^{*})}d{\bf r}^{*}+\frac{1}{8}\int \rho({\bf r}^{*})I({\bf r}^{*})d{\bf r}^{*}+(N-1)\int\rho({\bf r}^{*})C({\bf r}^{*})d{\bf r}^{*}.
\label{fhk}
\end{equation}
As a simple consistency check of our proposed approach, we consider the example
of the homogeneous interacting electron gas. In this case the total Hamiltonian is
$K+V_{ee}$, i.e. there is no external potential  $v({\bf r})$, thus, applying
the procedure yields to the Local Density Approximation (LDA) approximation to
the functional \cite{py}.
\section{Practical Utility}
In previous work it has already appeared the idea of employing the PI approach within the framework of DFT \cite{wy1,wy2}, but the main aim
there was avoiding the use of orbitals within the Kohn-Sham approach where
an exchange and correlation functional, $E_{xc}[\rho]$, was predefined.
The intention of this work, instead, is that of describing a procedure, rigorous
from the conceptual and numerical point of view, to make it possible the
numerical calculation of the \textit{exact} energy density for a given system (external potential), and thus use this information for developing analytic functionals. 
The advantage of the approach used here is that the energy density we derive is rigorously divided in its kinetic $I({\bf r})$ and potential $C({\bf r})$ components and thus the physical interpretation emerges in a natural way.\\
In practical terms, a possible way to use
this procedure is to treat basic reference systems (e.g., single atoms, small molecules) 
for which the application of the GSPI approach is computationally feasible.
This would allow for the determination of a database of energy densities that can be used
for the development of energy functionals, and to have a novel insight into
the basic physics of the functional in terms of each of its specific components. For instance, an accurate DFT-level description of the van der Waals interactions, with current functionals, for a system as simple as the helium dimer, is still an open problem \cite{springall}. 
Indeed, QMC calculations for the helium dimer are carried on to have some understanding of such interactions with the intention of using the results to build better functionals on a sound physical basis (see eg. Refs.\cite{springall, wux} and references therein). The approach suggested here would not be computationally more demanding than that of the QMC calculations of Refs.\cite{springall, wux}, however it would automatically provide the detailed (i.e. of each of the energy components) physics of the energy density of the ground state and thus a numerical reference in the development of energy functionals.
In particular, for a given system, one may treat
the problem for the case of interacting and for the case of not interacting
electrons and calculate $I_{int}({\bf r})$ and $C({\bf r})$ for
the interacting case, and $I_{nint}({\bf r})$ for the non interacting case. In this way one can determine $\epsilon_{xc}({\bf r})=I_{int}({\bf r})+(N-1)C({\bf r})-I_{nint}({\bf r})-\int\frac{\rho({\bf r}^{\prime})}{|{\bf r}-{\bf r}^{\prime}|}d{\bf r}^{\prime}$, (where$\int\frac{\rho({\bf r}^{\prime})}{|{\bf r}-{\bf r}^{\prime}|}d{\bf r}^{\prime}$ is
the Hartree term), that is the exchange and correlation energy density per particle. This quantity can be used as a basis for developing more general expressions of $\epsilon_{xc}({\bf r})$. Moreover, a newly developed functional  must give the $\epsilon_{xc}({\bf r})$ obtained by the procedure proposed here, when used for calculating the ground state of a specific system.  
In this context, relevant, long standing questions as that of the kinetic contribution to the $E_{xc}$ \cite{higuchi} or the problem of how to extend the LL formulation to the kinetic energy density \cite{ypl} could be now addressed in a more robust way. 
From the numerical point of view, the major advantage is that every time a
GSPI QMC calculation is done for a given system, the quantities $I({\bf r})$
and $C({\bf r})$ can be automatically determined at no additional cost. This
means that one would automatically produce an increasingly larger database to
be used for the development of functionals, and thus no more restricted to the
uniform gas only.
Furthermore, in this context, recently developed approaches within the framework of the so-called kernel-based Machine Learning (ML), propose a training strategy for
systematically determine a numerical functional \cite{ML} (though for the moment only for a simple proof-of-concept test case). ML
is a powerful tool for finding patterns in high-dimensional data and, applied to our strategy, would mean using GSPI exact
results for non trivial cases to refine (train) a numerical expression of the functional. It is remarkable that the flexibility of ML
allows for the insertion in the functional form of as much physical intuition as felt necessary (e.g., by imposing known exact
analytic constraints), in order to reduce the dimensionality parameter space in which the numerical optimization is performed.

\section{Warnings}
It must be noticed, that the use of $f$ in  QMC procedure, could lead to results characterized by a large variance. In principle, one may derive $\epsilon({\bf r})$ of Eq.\ref{deff1} in more efficient manners without passing through the calculation of $f$, however the separation of the kinetic functional in the two terms: $\frac{|\nabla\rho({\bf r}|^{2}}{\rho({\bf r})}$; $I({\bf r})$; would be no more straightforward and thus the detailed understanding of the physics related to the interplay between these two terms may be lost. While this is not relevant for practical applications, it may be relevant for the understanding of the basic physics and thus for the construction of analytic functionals. In general, without invoking $f$, in the standard GSPI procedure, $C({\bf r})$ can be efficiently calculated in the following way: the density is determined as the number of electron visiting some volume elements from the middle time slice, it follows that $C({\bf r})$ is the average electron-electron potential in that volume elements. However, it is not clear how one must then deal with the kinetic part (which is of major concern in this paper) although, as underlined above, it cannot be excluded that it may exists a more efficient way that does not directly involve $f$.
In case such a procedure is possible, our basic idea of determining each term of the energy density via GSPI QMC remains valid and we would gain in computational efficiency. At the current stage this aspect goes beyond the aim of this paper.    
\section{Universal Functional}
In the procedure discussed in the previous sections, we must restrict the
calculations to cases where $v({\bf r})=V_{ne}({\bf r})$, is specified and
then explicitly used in the QMC calculation of the transitional probabilities:
\begin{equation}
\left<{\bf R}_{i}|e^{-\tau V}|{\bf R}_{i+1}\right>=\frac{1}{(2\pi\tau)^{3N/2}}e^{-\frac{\tau}{2}[V_{ee}({\bf R}_{i})+V_{ne}({\bf R}_{i})+V_{ee}({\bf R}_{i+1})+V_{ne}({\bf R}_{i+1})]}.
\label{pot2}
\end{equation}
For this reason one cannot determine $F[\rho]$ as a universal functional of
$\rho$ independently from $v({\bf r})$. Here we propose to modify the GSPI
approach so that the resulting $f$ is a functional of $\rho$ only,
independently from $v({\bf r})$.\\
To this aim we rewrite the transitional probability for the potential part as:
\begin{equation}
\left<{\bf R}_{i}|e^{-\tau V}|{\bf R}_{i+1}\right>=\frac{1}{(2\pi\tau)^{3N/2}}e^{-\frac{\tau}{2}[V_{ee}({\bf R}_{i})+V_{ee}({\bf R}_{i+1})]}
\label{pot3}
\end{equation}
i.e., considering only the electron-electron interaction. The transitional probability for the kinetic part remains the same as above (Eq. \ref{spring}).
Next, we can calculate $f$, and thus $I({\bf r})$ and $C({\bf r})$ as in Eq.\ref{effe1},
Eq.\ref{irho} and Eq.\ref{crho}, but with a sampling restricted to a trial
$\rho_{trial}$. This means sampling the ${\bf R}_{i}'s$ in the configuration space
with the constrain that the one particle density is $\rho_{trial}$, for more technical details about sampling at given density see note in Ref.\cite{note1}. 
In this case the QMC procedure assures that the principle:
$\min_{f}\Gamma[\rho,f]$, is achieved in the sense that the resulting $\Gamma$
is that of {\it  ``ground state''} at the fixed $\rho_{trial}$.
Note that at this stage the external potential is not invoked and it is actually absent; in practice what we have done is to find the $f$ of ground state of a gas of electron with some {\it artificially forced} electron density. 
From the obtained $f$ we can now calculate the corresponding $I({\bf r})=I_{\rho_{trial}}({\bf r})$ and $C({\bf
  r})=C_{\rho_{trial}}({\bf r})$. These quantities are taken as a first guess to write an energy functional for a generic $\rho({\bf r})$:
\begin{equation}
E[\rho]=\int\rho({\bf r})\left[\frac{1}{8}\frac{|\nabla\rho({\bf r})|^{2}}{\rho({\bf r})^{2}}d{\bf
  r}+\frac{1}{8}I_{\rho_{trial}}({\bf r})+(N-1)C_{\rho_{trial}}({\bf r})+v({\bf
    r})\right]d{\bf r}.
\label{funtrial}
\end{equation}
Next, we use Eq.\ref{funtrial} for a minimization
w.r.t. $\rho$ and obtain a new $\rho=\rho^{1}_{out}$, different from
$\rho_{trial}$ because in Eq.\ref{funtrial} the effect of $v({\bf r})$ is explicitly included during the energy minimization.
At this point one can use $\rho^{1}_{out}$ as a new trial density, repeat the 
QMC procedure, that is we search the ground state of a gas of electrons with 
{\it artificially forced} electron density $\rho^{1}_{out}$, this will lead to a new $f$ and in turn to a new $I({\bf r})$ and $C({\bf r})$ and thus we can have a new guess for $E[\rho]:\int\rho({\bf r})\left[\frac{1}{8}\frac{|\nabla\rho({\bf r})|^{2}}{\rho({\bf r})^{2}}d{\bf
  r}+\frac{1}{8}I_{\rho^{1}_{out}}({\bf r})+(N-1)C_{\rho^{1}_{out}}({\bf r})+v({\bf
    r})\right]d{\bf r}.$ As above we can then use the expression $E[\rho]$ (again for a generic $\rho({\bf r})$) for a minimization w.r.t. $\rho({\bf r})$ and obtain as a result a new $\rho^{2}_{out}$ and repeat the procedure until $\rho^{i}_{out}=\rho^{i+1}_{out}$ with some accuracy.
Of course, convergence must be proven and intuition suggests to start from some ``reasonable'' $\rho_{trial}$.
However, beyond the several problems that a ``realistic implementation'' would imply, this procedure would have at least some conceptual benefits; this is a real space procedure which does not require neither orbitals nor a predefinition of
the functional as it is instead the case for the Kohn-Sham approach, but above all, the 
procedure has the potential to deliver the energy functional on-the-fly during the
calculation. Since this approach is valid for any system, independently from
$v({\bf r})$, this is an implicit way to define in iterative manner the
universal functional of Hohenberg and Kohn. Despite the
involved computational costs are not clear yet, this idea may in principle offer
a complementary way for using QMC in the perspective of DFT and perhaps a path to find a compromise between the high accuracy and computational costs of QMC and the low accuracy and computational costs of DFT. In any case the optimization of the computational aspects of this idea
represent an interesting challenge for future research.

\section{Conclusions}
We have proposed a theoretical/conceptual protocol based on using the Ground State Path Integral Quantum Monte
Carlo technique in the context of DFT for the
determination of the energy functional. We have shown that the method can be
certainly used to calculate automatically the energy density of the ground
state in terms similar to those of the energy functional in DFT. This allows for using
the results for building a database for the development and control of energy functionals beyond the
standard case of the uniform electron gas as a reference. A second
possibility to employ GSPI QMC is that of using the procedure as an intermediate step in an iterative loop to determine the density of ground state within the Levy-Lieb energy functional minimization procedure: In simple terms the functional is determined on-the-fly during the minimization procedure.
Being valid for any external potential, this procedure implicitly defines, in
numerical terms, the universal functional of Hohenberg and Kohn. Despite the fact that the
computational costs may turn out to be rather high, the procedure may open a way to find a compromise between the accuracy of
QMC and the feasibility of DFT. As underlined above, next challenge would be that of searching for the most efficient computational implementation of the idea and compare its performance with that of standard methods. However, even if it will turn out to be computationally less convenient, this protocol would always assure the access to basic physical information that can be then employed as a complementary knowledge in the development of new, physically sound, energy functional; this in summary is the message of this paper.
\begin{acknowledgements} 
 We thank Kieron Burke and Matthias Scheffler for a critical reading of the manuscript. This work was partially supported by the Heisenberg Stipendium of the Deutsche Forschungsgemeinschaft (DFG) (grant code DE 1140/5-1) provided to L.D.S
\end{acknowledgements}

\section*{Appendix}
Here we report the analytic calculations of $\nabla_{{\bf r}^{*}}[f({\bf r}^{*}_{2},....{\bf r}^{*}_{N}|{\bf r}^{*})]$.
Let us define 
\begin{equation}
f({\bf r}^{*}_{2},....{\bf r}^{*}_{N}|{\bf r}^{*})=\frac{1}{Z_{{\bf r}^{*}}}\times W({\bf r}^{*},{\bf r}^{*}_{2},...{\bf r}^{*}_{N})
\end{equation} 
with 
\begin{eqnarray}
 W({\bf r}^{*},{\bf r}^{*}_{2},...{\bf r}^{*}_{N})=\int d{\bf R}_{0}d{\bf R}_{1}.......d{\bf R}_{\frac{M}{2}-1}d{\bf R}_{\frac{M}{2}+1}.......d{\bf R}_{M} \nonumber\\ \psi({\bf R}_{0})\exp[-S({\bf R}_{*}, {\bf
    R}_{0},{\bf R}_{1},....{\bf R}_{\frac{M}{2}-1},{\bf R}_{\frac{M}{2}+1}......{\bf R}_{M})]\psi({\bf R}_{M})
\label{doppiaw}
\end{eqnarray}
It follows that:
\begin{equation}
\nabla_{{\bf r}^{*}}[f({\bf r}^{*}_{2},....{\bf r}^{*}_{N}|{\bf r}^{*})]=-\frac{1}{(Z_{{\bf r}^{*}})^{2}}W({\bf r}^{*},{\bf r}^{*}_{2},...{\bf r}^{*}_{N})\times \nabla_{{\bf r}^{*}}Z_{{\bf r}^{*}}+\frac{1}{Z_{{\bf r}^{*}}}\times\nabla_{{\bf r}^{*}}W({\bf r}^{*},{\bf r}^{*}_{2},...{\bf r}^{*}_{N})
\label{st1}
\end{equation}
with
\begin{eqnarray}
 \nabla_{{\bf r}^{*}}Z_{{\bf r}^{*}}=\int d{\bf R}_{*}^{N-1}d{\bf R}_{0}d{\bf R}_{1}.......d{\bf R}_{\frac{M}{2}-1}d{\bf R}_{\frac{M}{2}+1}.......d{\bf R}_{M} \nonumber\\ \psi({\bf R}_{0})\nabla_{{\bf r}^{*}}\left(\exp[-S({\bf R}_{*}, {\bf
    R}_{0},{\bf R}_{1},....{\bf R}_{\frac{M}{2}-1},{\bf R}_{\frac{M}{2}+1}......{\bf R}_{M})]\right)\psi({\bf R}_{M}).
\label{nablaz}
\end{eqnarray}
In, $\nabla_{{\bf r}^{*}}\left(\exp[-S({\bf R}_{*}, {\bf
    R}_{0},{\bf R}_{1},....{\bf R}_{\frac{M}{2}-1},{\bf
    R}_{\frac{M}{2}+1}......{\bf R}_{M})]\right)$, the terms which are involved
into the derivation are only those with ${\bf R}_{*}$. This means that the
interesting quantity to calculate is: $\nabla_{{\bf r}^{*}}\left[\left<{\bf
    R}_{\frac{M}{2}-1}|e^{-\tau H}|{\bf R}_{*}\right>\left<{\bf
    R}_{*}|e^{-\tau H}|{\bf R}_{\frac{M}{2}+1}\right>\right]$, the other terms
of $\exp[-S]$ factorize.
This gives:
\begin{eqnarray}
\nabla_{{\bf r}^{*}}\left[\left<{\bf R}_{\frac{M}{2}-1}|e^{-\tau H}|{\bf
    R}_{*}\right>\left<{\bf R}_{*}|e^{-\tau H}|{\bf
    R}_{\frac{M}{2}+1}\right>\right]=\nonumber\\\nabla_{{\bf r}^{*}}\left[\frac{1}{(2\pi\tau)^{3N/2}}e^{-\frac{\tau}{2}\left(\frac{{\bf R}_{\frac{M}{2}-1}-{\bf R}_{*}}{\tau}\right)^{2}}e^{-\frac{\tau}{2}\left(\frac{{\bf R}_{*}-{\bf R}_{\frac{M}{2}+1}}{\tau}\right)^{2}}e^{-\frac{\tau}{2}[V({\bf R}_{\frac{M}{2}-1})+V({\bf R}_{*})]}
e^{-\frac{\tau}{2}[V({\bf R}_{*})+V({\bf R}_{\frac{M}{2}+1})]}\right].
\label{exp2}
\end{eqnarray}
For the kinetic part one has:
\begin{eqnarray}
\nabla_{{\bf r}^{*}}\left[\frac{1}{(2\pi\tau)^{3N/2}}e^{-\frac{\tau}{2}\left(\frac{{\bf R}_{\frac{M}{2}-1}-{\bf R}_{*}}{\tau}\right)^{2}}e^{-\frac{\tau}{2}\left(\frac{{\bf R}_{*}-{\bf R}_{\frac{M}{2}+1}}{\tau}\right)^{2}}\right]=-\left[\frac{1}{(2\pi\tau)^{3N/2}}e^{-\frac{\tau}{2}\left(\frac{{\bf R}_{\frac{M}{2}-1}-{\bf R}_{*}}{\tau}\right)^{2}}e^{-\frac{\tau}{2}\left(\frac{{\bf R}_{*}-{\bf R}_{\frac{M}{2}+1}}{\tau}\right)^{2}}\right]\times\nonumber\\ \tau[({\bf r}^{\frac{M}{2}-1}-{\bf r}^{*})-({\bf r}^{*}-{\bf r}^{\frac{M}{2}+1})]
\label{kpart}
\end{eqnarray}
that is:
\begin{equation}
\nabla_{{\bf r}^{*}}\left[\frac{1}{(2\pi\tau)^{3N/2}}e^{-\frac{\tau}{2}\left(\frac{{\bf R}_{\frac{M}{2}-1}-{\bf R}_{*}}{\tau}\right)^{2}}e^{-\frac{\tau}{2}\left(\frac{{\bf R}_{*}-{\bf R}_{\frac{M}{2}+1}}{\tau}\right)^{2}}\right]=\tau({\bf r}^{\frac{M}{2}+1}-{\bf r}^{\frac{M}{2}-1})\times \left[\frac{1}{(2\pi\tau)^{3N/2}}e^{-\frac{\tau}{2}\left(\frac{{\bf R}_{\frac{M}{2}-1}-{\bf R}_{*}}{\tau}\right)^{2}}e^{-\frac{\tau}{2}\left(\frac{{\bf R}_{*}-{\bf R}_{\frac{M}{2}+1}}{\tau}\right)^{2}}\right].
\label{kpart2}
\end{equation}
For the potential part:
\begin{eqnarray}
\nabla_{{\bf
    r}^{*}}\left[\frac{1}{(2\pi\tau)^{3N/2}}e^{-\frac{\tau}{2}\left(V({\bf
      R}_{*})+V({\bf R}_{\frac{M}{2}-1})\right)}e^{-\frac{\tau}{2}\left(V({\bf
      R}_{*})+V({\bf R}_{\frac{M}{2}+1})\right)}\right]=-\tau\nabla_{{\bf
      r}^{*}}V({\bf
    R}_{*})\times\nonumber\\\times\left[\frac{1}{(2\pi\tau)^{3N/2}}e^{-\frac{\tau}{2}\left(V({\bf
        R}_{*})+V({\bf R}_{\frac{M}{2}+1})\right)}e^{-\frac{\tau}{2}\left(V({\bf
      R}_{*})+V({\bf R}_{\frac{M}{2}-1})\right)}\right].
\label{potpart}
\end{eqnarray}
It follows:
\begin{eqnarray}
\nabla_{{\bf r}^{*}}Z_{{\bf r}^{*}}=\int d{\bf R}^{N-1}_{*}d{\bf R}_{0}d{\bf
  R}_{1}....d{\bf R}_{\frac{M}{2}-1}d{\bf R}_{\frac{M}{2}+1}......d{\bf
  R}_{M}\left[({\bf r}^{\frac{M}{2}+1}-{\bf
    r}^{\frac{M}{2}-1})-\tau\nabla_{{\bf r}^{*}}V({\bf R}_{*})\right]\times
\nonumber\\ \times\psi({\bf R}_{0})\exp[-S({\bf R}_{*}, {\bf
    R}_{0},{\bf R}_{1},....{\bf R}_{\frac{M}{2}-1},{\bf
    R}_{\frac{M}{2}+1}......{\bf R}_{M})]\psi({\bf R}_{M})
\label{finz}
\end{eqnarray}
It follows also that:
\begin{eqnarray}
 \nabla_{{\bf r}^{*}}W({\bf r}^{*},{\bf r}^{*}_{2},...{\bf r}^{*}_{N})=\int d{\bf R}_{0}d{\bf
  R}_{1}....d{\bf R}_{\frac{M}{2}-1}d{\bf R}_{\frac{M}{2}+1}......d{\bf
  R}_{M}\left[({\bf r}^{\frac{M}{2}+1}-{\bf
    r}^{\frac{M}{2}-1})-\tau\nabla_{{\bf r}^{*}}V({\bf R}_{*})\right]\times
\nonumber\\ \times\psi({\bf R}_{0})\exp[-S({\bf R}_{*}, {\bf
    R}_{0},{\bf R}_{1},....{\bf R}_{\frac{M}{2}-1},{\bf
    R}_{\frac{M}{2}+1}......{\bf R}_{M})]\psi({\bf R}_{M})
\label{finw}
\end{eqnarray}
The term $({\bf r}^{\frac{M}{2}+1}-{\bf
    r}^{\frac{M}{2}-1})$ can be calculated without any additional
computational cost during the sampling, however this is true also for the term
$\nabla_{{\bf r}^{*}}V({\bf R}_{*})$. In fact the explicit form of $V({\bf R}_{*})$ is known
and thus its gradient can be calculated analytically and sampled without
additional computational costs.
If one uses the expression of Eq.\ref{finz} and Eq.\ref{finw} into $I({\bf
  r}^{*})$  ($C({\bf r}^{*})$ is straightforward) obtains the exact form of
the Hohenberg-Kohn functional.

\end{document}